\documentclass[a4paper]{article}
\usepackage{graphicx}
\begin{document}

\title{Conduction Effect of Thermal Radiation in a Metal Shield Pipe in a Cryostat for a Cryogenic Interferometric Gravitational Wave Detector}

\author{
Takayuki Tomaru$^A$, Masao Tokunari$^B$, Kazuaki Kuroda$^B$, \\
Takashi Uchiyama$^B$, Akira Okutomi$^B$, Masatake Ohashi$^B$, \\
Hiroyuki Kirihara$^B$, Nobuhiro Kimura$^A$, Yoshio Saito$^A$, \\
Nobuaki Sato$^A$, Takakazu Shintomi$^C$, Toshikazu Suzuki$^A$, \\
Tomiyoshi Haruyama$^A$, Shinji Miyoki$^B$, Kazuhiro Yamamoto$^B$, \\
Akira Yamamoto$^A$ \\
\ \ \\
\it{\small{$^A$High Energy Accelerator Research Organization (KEK),}} \\
\it{\small{ 1-1 Oho, Tsukuba, Ibaraki 305-0801, Japan}} \\
\it{\small{$^B$Institute for Cosmic Ray Research, University of Tokyo, }} \\
\it{\small{5-1-5 Kashiwanoha, Kashiwa 277-8585, Japan}} \\
\it{\small{$^C$Nihon University, 4-2-1-602 Kudankita, Chiyoda 102-0073, Japan}}
}

\date{}

\maketitle

A large heat load caused by thermal radiation through a metal shield pipe was observed in a cooling test of a cryostat for a prototype of a cryogenic interferometric gravitational wave detector.
The heat load was approximately 1000 times larger than the value calculated by the Stefan-Boltzmann law.
We studied this phenomenon by simulation and experiment and found that it was caused by the conduction of thermal radiation in a metal shield pipe.

\section{Introduction}
A cryogenic interferometric gravitational wave detector is a next-generation detector for developing the field of gravitational wave astronomy.
One of the features of the detector is that it uses cryogenic mirrors in the optical interferometer to reduce thermal noises~\cite{1},
which are one of the principle sensitivity limitations.
At present, a Large-scale Cryogenic Gravitational wave Telescope (LCGT)~\cite{2} has been proposed in Japan.
As a prototype of the LCGT, the Cryogenic Laser Interferometer Observatory (CLIO)~\cite{3} has been constructed.
It has two 100\,m baselines and four cryogenic sapphire mirrors to form Fabry-Perot cavities.
Each sapphire mirror is set in an independent cryostat and cooled by ultra-low vibration cryocooler systems~\cite{4,5,6}.

\begin{figure}[htbp]
\begin{center}
\includegraphics[width=12cm,clip]{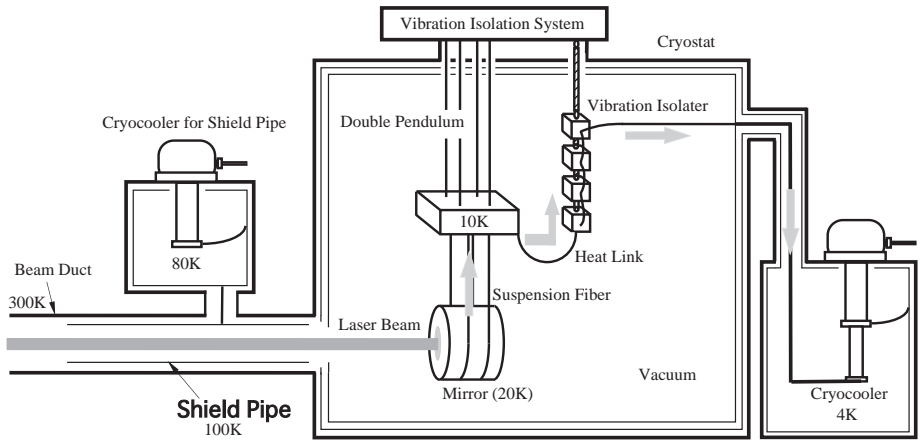}
\caption{Schematic representation of the CLIO cryostat. To reduce the solid angle between the 300\,K beam duct and the cryostat to reduce the thermal radiation heat load, a thermal radiation shield pipe with a diameter of 30\,cm and length of 5\,m was introduced in the beam duct. It was separated from the cryostat and independently cooled by an ultra-low 80\,K cryocooler system.}
\label{coolsys}
\end{center}
\end{figure}

Figure~\ref{coolsys} shows the schematic representation of the CLIO cryostat.
A feature of this cryostat is that it has a large laser opening that is facing a 300\,K beam duct since it is difficult to cool all of beam ducts in huge interferometers.
In the case of CLIO, the diameter of the laser opening is 30\,cm.
To reduce the solid angle between the 300\,K area and the cryostat, i.e., to reduce the thermal radiation heat load from the 300\,K area to the cryogenic area, 
an aluminum pipe was installed as a thermal radiation shield in the beam duct.
The diameter and length of the pipe are 30\,cm and 5\,m, respectively.
It is separated from the main cryostat and independently cooled to 100\,K by a vibration-free 80\,K cryocooler system.

In general, while designing cryostats, the thermal radiation heat load from an area with a temperature $T_2$ to an area with a temperature $T_1$ was estimated by using the Stefan-Boltzmann law:
\begin{equation}
\label{eq1}
 P = \frac{\epsilon_1 \epsilon_2}{\epsilon_1+\epsilon_2 - \epsilon_1 \epsilon_2} \sigma (T_2^4 - T_1^4) A \frac{\Omega}{2 \pi},
\end{equation}
where $\epsilon_i$ ($i = 1,2$) is the emissivity; $\sigma$, the Stefan-Boltzmann constant; $A$, the surface area; and $\Omega$, the solid angle between both the areas.
This equation implies that the heat generated by thermal radiation is transferred from one surface to the other.
The thermal radiation heat load in the CLIO cryostat was also estimated using this equation; 
however, an extremely large heat load over 3\,W, which was approximately 1000 times larger than the design, was found in its cooling test.
Since the cryostat reached the design temperature when the laser opening was closed, the large heat load could be generated due to the thermal radiation through the shield pipe.

Here, it should be noted that 300\,K black body radiation consists of infrared (IR) light with a peak wavelength of $20\,\mathrm{\mu m}$.
Moreover, the IR reflectivity of aluminum is sufficiently large.
Therefore, thermal radiation could be reflected and conducted through an aluminum shield pipe.
We investigated this phenomenon by simulation and experiment.

\section{Analysis by Simulation} \label{sim}

\begin{figure}[htbp]
\begin{center}
\includegraphics[width=12cm,clip]{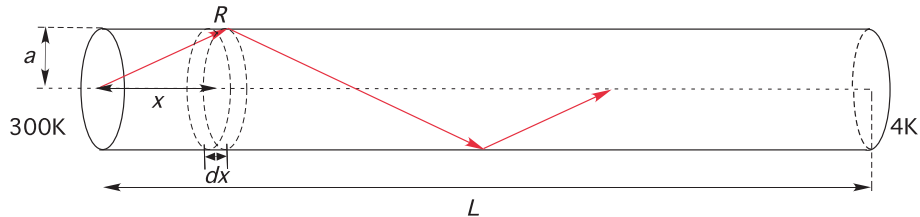}
\caption{Ray-trace model for conduction of thermal radiation in a cylindrical pipe.}
\label{model1}
\end{center}
\end{figure}

To investigate the hypothesis of the conduction phenomenon of thermal radiation in a radiation shield pipe in the CLIO, we simulated it by using a ray-trace model.
Figure~\ref{model1} shows the schematic representation of the model.
When IR light is isotropically emitted from the 300\,K area, the power incident onto the area $dx$ at a distance $x$ from the end of the pipe is given as follows:
\begin{eqnarray}
\label{eq2}
 P(x) = P_0 \frac{d\Omega}{2 \pi}, \\
 \label{eq3}
 d\Omega = \frac{2 \pi a^2\,dx}{(x^2+a^2)^{3/2}},
\end{eqnarray}
where $P_0$ is the total emitted power from the 300\,K area; $d\Omega$, the solid angle; and $a$, the pipe radius.
This IR light reflects $N(x)$ times inside the pipe wall with a reflectivity of $R$.
When the length of the pipe is $L$, the number of reflections $N(x)$ is given by
\begin{equation}
\label{eq4}
 N(x) = 1 + n\Bigr(\frac{L-x}{2 x}\Bigr),
\end{equation}
where $n((L-x)/2 x)$ is a function that $(L-x)/2 x$ is converted to integer.
Although the reflectivity $R$ depends on the wavelength and incident angle of light, we assumed it to be constant in this model.
From the above equations, the IR power $P_{ref}$ reaching the 4\,K area after reflection in the pipe is calculated as
\begin{equation}
\label{eq5}
 P_{ref} = P_0 \int_{0}^{L} R^{N(x)} \frac{d\Omega}{2 \pi} = P_0 \int_{0}^{L} R^{N(x)} \frac{a^2}{(x^2+a^2)^{3/2}}\,dx .
\end{equation}

On the other hand, the direct power incident on the 4\,K area without reflection, which is described by the Stefan-Boltzmann law in equation~(\ref{eq1}), is rewritten as
\begin{equation}
\label{eq6}
 P_{SB} = P_0 \frac{a^2}{2 L^2}.
\end{equation}

\begin{table}[htdp]
\caption{Simulation result of conduction effect of thermal radiation by a IR ray-trace model in the case of the CLIO. 
$R$ shows IR reflectivity of aluminum pipe. $(P_{ref} + P_{SB})/(P_{SB})$ implies the ratio of the total heat load through the shield pipe to the direct incident power without reflection, and $P_{ref}/P_0$ implies the ratio of the heat load with reflection in the pipe to the total emitted heat power from 300\,K.}
\begin{center}
\begin{tabular}{|c|c|c|c|}
\hline
\ \ & $R = 0.90$ & $R = 0.95$ & $R = 0.97$ \\
\hline
$(P_{ref} + P_{SB})/(P_{SB})$ & 307 & 622 & 898 \\
\hline
$P_{ref}/P_0$ &14\% & 28\% & 40\% \\
\hline
\end{tabular}
\end{center}
\label{tab1}
\end{table}%

Table~\ref{tab1} lists the calculated results.
This result shows that the total power reaching the 4\,K area through a shield pipe was several hundreds times larger than the direct power incident without reflection. 
The conduction effect accounted for several dozen percent of the total emitted power from the 300\,K area.
Therefore, the simulation result supported the hypothesis that the large heat load obtained in the CLIO cryostat was possibly caused by the conduction effect of thermal radiation in the radiation shield pipe.

\section{Experimental Verification}

\begin{figure}[htbp]
\begin{center}
\includegraphics[width=12cm,clip]{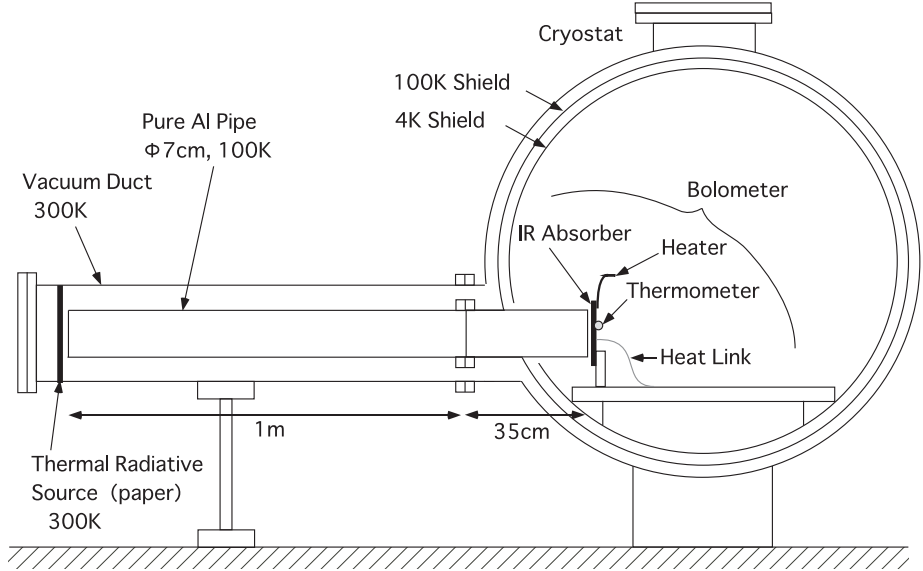}
\caption{Experimental setup for measurement of heat load through an aluminum shield pipe by using a cryostat of early prototype of cryogenic interferometer, called CLIK. The method for measurement was calorimetry.}
\label{setup}
\end{center}
\end{figure}

To verify the phenomenon of thermal radiation conduction in the shield pipe, we performed a model experiment by using a cryostat of the early prototype of a cryogenic interferometer, called CLIK.
Figure~\ref{setup} shows the experimental setup.
We used the calorimetric method.
A pure aluminum pipe with a diameter of 7\,cm, length of 1.35\,m, and temperature of 100\,K, was installed in a vacuum duct.
A thermal radiative source was set in front of the 300\,K side of the pipe, 
and a bolometer was placed at the opposite side of the pipe in the cryostat.
Paper was used as the thermal radiative source since it is well known that it has a large emissivity of approximately 0.9.
The bolometer consisted of an IR absorber called ultra-black NiP (UB-NiP), a film heater, a thermometer, and a thin aluminum heat link.
The UB-NiP had a large IR absorption rate of approximately 80\% and small outgassing ratio in a high vacuum system~\cite{7}.
A silicon thermometer made by Lake Shore Inc. was attached to the UB-NiP in order to measure its temperature.
A film heater was used to calibrate the relation between the heat load and the temperature of the bolometer, 
and the heat link was a heat path between the bolometer and the 4\,K stage in the cryostat.

The measurement was performed twice and good repeatability within 2\% was confirmed.
From this measurement, we observed a heat load of 390\,mW on an average.
This value was compensated by using the emissivity of paper and absorption rate of the UB-NiP.
This observed heat load was 740 times larger than the value estimated by the Stefan-Boltzmann law in equation~(\ref{eq6}).

To confirm that the experimentally observed heat load was certainly caused by the conduction effect of thermal radiation modeled in section~\ref{sim}, 
we checked the IR reflectivity of the aluminum shield pipe by using an IR free electron laser (FEL) at Tokyo University of Science~\cite{8}.
In this measurement, the regularly reflected laser power on a sample surface was compared to the injected laser power.
The sample was fabricated by cutting a part of the aluminum shield pipe used in the CLIK experiment. 
Although large errors of several percent due to the instability of the FEL power and inaccuracy of detectors were present in the measurement data, the reflectivity of the aluminum pipe at a wavelength of $8.33\,\mu\mathrm{m}$ was approximately 98\%, and its dependence on the laser injection angle was small.
The IR reflectivity of the aluminum pipe estimated from the CLIK experiment and equation~(\ref{eq5}) was 95.0\%.
Therefore, we concluded that these results showed the legitimacy of the conduction model of thermal radiation in a shield pipe.

\section{Discussion}

\begin{figure}[htbp]
\begin{center}
\includegraphics[width=8cm,clip]{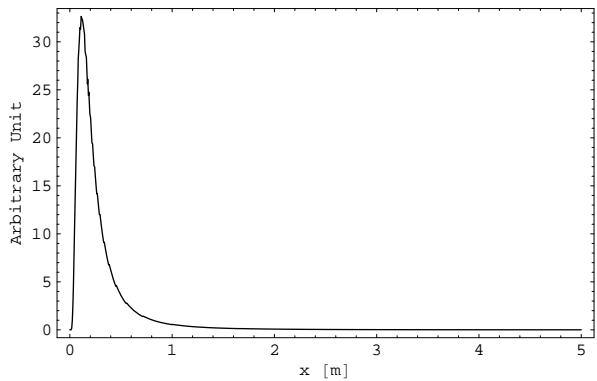}
\caption{Plot of integrand in equation~(\ref{eq5}) in the case of the CLIO. The unit of the vertical axis is arbitrary. The origin of the horizontal axis ($x$) was set as the end of the aluminum pipe of the 300\,K side.}
\label{integrand}
\end{center}
\end{figure}

The reduction in the conduction effect of thermal radiation is an important issue for the cryostat design of cryogenic interferometers.
In a simple case in which a part of the shield pipe has a small IR reflectivity, we can easily estimate its reduction rate by using the ray-trace model described in section~\ref{sim}.

Figure~\ref{integrand} shows a plot of the integrand in equation~(\ref{eq5}) in the case of the CLIO.
This graph shows that the incident power at $x<1\,\mathrm{m}$ mostly affects the heat load in the cryostat.
Therefore, it is effective to introduce an IR absorption area at the 300\,K side of the pipe.
When a low reflectivity coating of $R'$ is used at a region between $x=0$ and $x=s$,
equation~(\ref{eq5}) is modified to
\begin{eqnarray}
\label{eq7}
P_{ref}' = P_0\Big[ \int_{0}^{s} R'^{N(x,s)} R^{N(x,L)-N(x,s)} \frac{a^2}{(x^2+a^2)^{3/2}}\,dx \nonumber \\
+ \int_{s}^{L} R^{N(x,L)} \frac{a^2}{(x^2+a^2)^{3/2}}\,dx \Big] .
\end{eqnarray}

\begin{table}[htdp]
\caption{Calculation results of $(P_{ref}' + P_{SB})/P_{SB}$ in equation~(\ref{eq7}). $(P_{ref}' + P_{SB})/(P_{SB})$ implies the ratio of the total heat load through the shield pipe to the direct incident power without reflection, and $(P_{ref}' + P_{SB})/(P_{ref} + P_{SB})$ implies the reduction rate of the total heat load. In this calculation, the reflectivity $R$ of aluminum pipe and reflectivity $R'$ of absorption area were assumed to be 95.0\% and 10.0\%, respectively.}
\begin{center}
\begin{tabular}{|c|c|c|c|}
\hline
$s$ [m] & 0.5 & 1.0 & 2.5 \\
\hline
$(P_{ref}' + P_{SB})/P_{SB}$ & 111 & 38 & 7.4 \\
\hline
$(P_{ref}' + P_{SB})/(P_{ref} + P_{SB})$ & 18\% & 6\% & 1\% \\
\hline
\end{tabular}
\end{center}
\label{tab2}
\end{table}%

Table~\ref{tab2} shows the calculated results of $(P_{ref}' + P_{SB})/P_{SB}$.
This result indicated that the IR absorption area was required for more than half of the shield pipe.

\section{Conclusion}
We studied the conduction effect of thermal radiation in a cryogenic shield pipe.
Simulation results showed that thermal radiation could be reflected and conducted through the shield pipe and that the heat load caused by this effect could be several hundred times larger than the estimated value by the Stefan-Boltzmann law.
A model experiment showed that the observed heat load was 740 times larger than the direct incident power calculated by the Stefan-Boltzmann law.
From these results, a 95.0\% IR reflectivity of the cryogenic shield pipe was derived in the CLIK experiment, and this value was consistent with the result of the IR reflectivity measurement of the pipe by using the IR laser.
Therefore, we confirmed the occurrence of the conduction effect of thermal radiation in a long shield pipe.
Although the reduction of this effect was briefly discussed, a precise study of its reduction through experiments is a future issue.

\section*{Acknowledgements}
We would like to extend our thanks to Dr.\ T. Imai and Prof. K. Tsukiyama of the IR FEL Research Center, Tokyo University of Science, for the measurement of the IR reflectivity of an aluminum pipe.
This work is partly supported by the cooperative program 
of the Institute for Cosmic Ray Research, University of Tokyo.

\end{document}